\documentclass[slac_one]{revtex4}
\usepackage{graphicx}
\usepackage{fancyhdr}
\begin{document}
{\hbox to\hsize {\hfill UFIFT-HEP-05-11 }}

\title{{\small{2005 International Linear Collider Workshop - Stanford,
U.S.A.}}\\ 
\vspace{12pt}
Testing Cosmology at the ILC} 

%

\author{A. Birkedal\footnote{This talk was given by A.~Birkedal, 
describing past and ongoing work performed in collaboration with the other authors.}, K. Matchev}
\affiliation{University of Florida, Gainesville, FL 32611, USA}
\author{J. Alexander, K. Ecklund, L. Fields, R. C. Gray, D. Hertz, C. D. Jones, J. Pivarski}
\affiliation{Cornell University, Ithaca, NY 14850, USA}

\begin{abstract}
We investigate the capabilities for the LHC and the ILC to perform measurements 
of new physics parameters relevant for the calculation of the cosmological relic 
abundance of the lightest neutralino in supersymmetry.  Specifically, we delineate
the range of values for the cold dark matter relic abundance $\Omega_{\chi} h^2$, 
which will be consistent with the expected precision measurements at the LHC, and,
subsequently, at the ILC. We illustrate our approach with a toy study of 
an ``updated benchmark'' point B'. We then show some
preliminary results of a similar analysis along those lines of the LCC2 benchmark 
point in the focus point region.
\end{abstract}

\maketitle

\thispagestyle{fancy}


\section{DARK MATTER AND SUPERSYMMETRY} 

By now there is overwhelming evidence for the existence of non-baryonic, non-luminous matter.  
The post-WMAP determination of the relic density $\Omega_\chi$ of this dark matter
is accurate at the level of 10\%~\cite{WMAP}:
\begin{equation}
0.094 \leq \Omega_{\chi} h^2 \leq 0.129 \ \  {\rm (at\ 2 \sigma)}.
\label{wmap}
\end{equation}
In spite of this astonishing accuracy, the precise nature of dark matter 
continues to remain one of the greatest unsolved mysteries in science.  
The preferred solution is to postulate the existence of a new stable, 
electrically neutral particle. One logical possibility 
is that this particle interacts only gravitationally with our world,
in which case very little can be said about its properties and furthermore,
any attempts to detect it directly and indirectly in the laboratory 
appear to be doomed. However, it is also possible that the dark matter
candidate has some additional interactions with the Standard Model
particles, and this possibility seems to be very well motivated by 
extensions of the Standard Model which strive to explain the origin
of electroweak symmetry breaking and the related gauge hierarchy problem.
These additional interactions could serve to keep the dark matter particles
in thermal equilibrium with the primordial soup in the early universe.
Given the strength of the dark matter interactions,
the standard freeze-out calculation~\cite{KT}
allows a prediction of the current relic abundance.
Conversely, the measurement of the amount of present-day dark matter (\ref{wmap})
determines the size of the total annihilation cross-section $\sigma_{\rm an}$ 
of dark matter~\cite{Birkedal:2004xn}:
$\sigma_{\rm an}\approx 0.85$ pb ($\sigma_{\rm an}\approx 7$ pb)
for dark matter particles annihilating in an $s$-wave ($p$-wave).
In a delicious empirical coincidence this is the typical size of the 
annihilation cross-section for a weakly interacting massive particle
(WIMP) with a mass $M_\chi$ in the range of $100\ {\rm GeV} - 1\ {\rm TeV}$.

Supersymmetric theories contain several such WIMP particles: the spin-1/2 
partners of the photon, $Z$ and neutral Higgs bosons~\cite{susy_review}.  
These states mix and their mass eigenstates are the neutralinos:
$\tilde\chi^0_i, i=1,2,3,4$. Among the myriad of supersymmetric models, 
those with a conserved quantum number called $R$-parity 
have attracted the most attention. $R$-parity guarantees proton stability
as well as a stable lightest superpartner (LSP).
In large regions of parameter space the LSP is the lightest neutralino 
$\tilde\chi^0_1$, which is an ideal WIMP dark matter candidate.

\section{DISCOVERING DARK MATTER AT COLLIDERS}

\begin{figure*}[t]
\centering
\includegraphics[width=80mm]{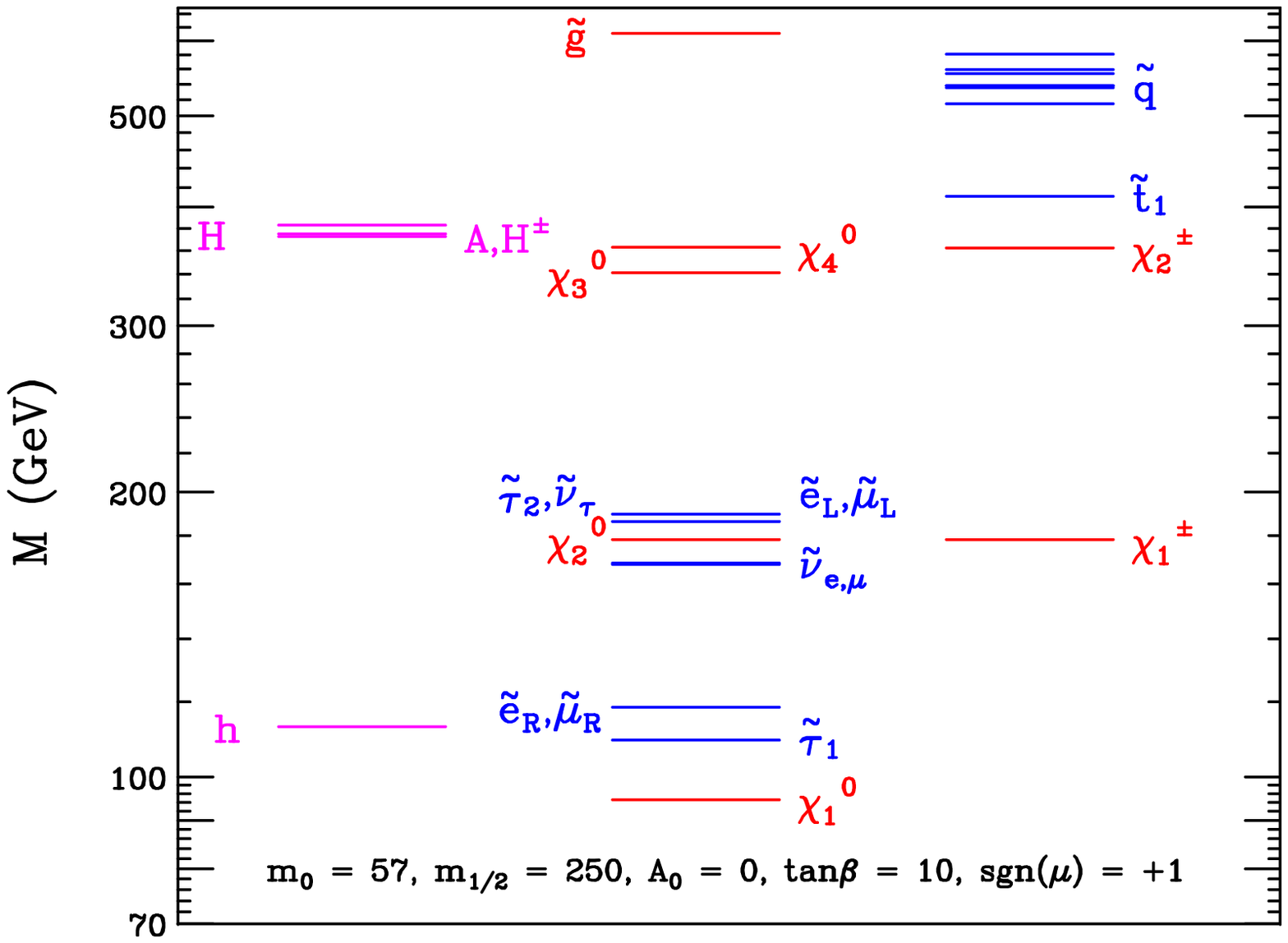}
\hspace{5mm}
\includegraphics[width=80mm]{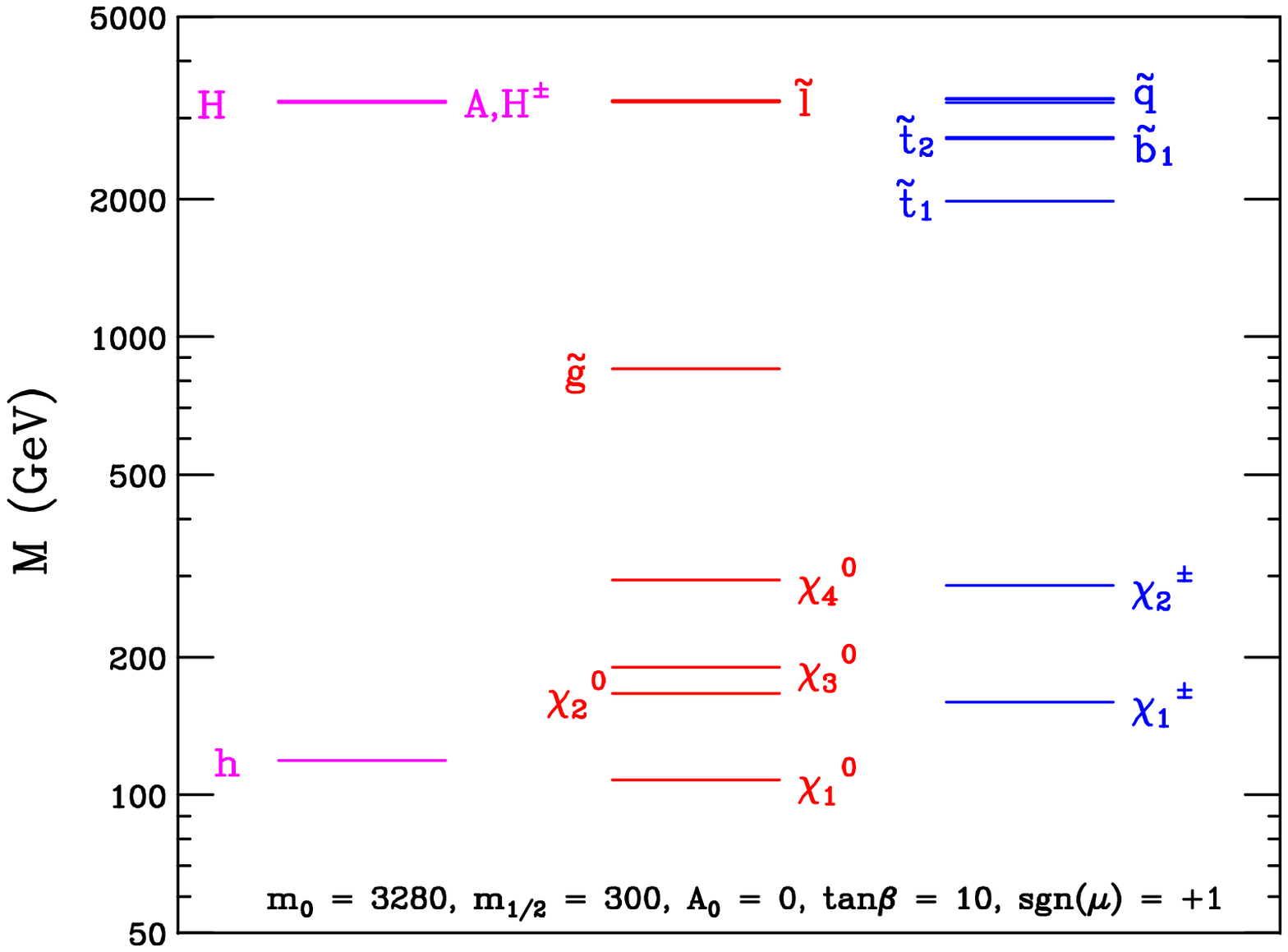}
\caption{Mass spectra of points B' (left) and LCC2 (right).}
\label{fig:spectra}
\end{figure*}

If the LSP {\em is} the dark matter particle, the generic collider signatures 
of supersymmetry (SUSY) all involve missing energy due to the two stable $\tilde\chi^0_1$s
escaping the detector. The observation of a missing energy signal at the LHC and/or ILC
will fuel the WIMP hypothesis. However, a missing energy signal at a collider 
only implies that particles have been created that are stable on a timescale 
characteristic of the detector size. In order to prove that the missing 
energy particle is indeed a viable WIMP dark matter candidate, one needs to calculate
its expected relic abundance today. To this end, one needs to measure all parameters 
(masses, couplings, mixing angles etc.) which enter the freeze-out calculation. 
In the most general minimal supersymmetric extension of the Standard Model (MSSM) 
there are more than 100 input parameters at the weak scale, but fortunately, a 
lot of them are either tightly constrained (e.g. $CP$-violating phases and flavor-violating
mixing angles) or not very relevant for the dark matter calculation. Nevertheless, 
there are still quite a number of relevant parameters left, which need to be
determined from collider data. Of course, the relevance of any one parameter
depends sensitively on the parameter space point. In this talk, as two illustrative
examples, we will consider an updated benchmark point B' \cite{bench}
and the LCC2 benchmark point \cite{Gray:2005ci}. The superpartner mass spectra 
for these points are illustrated in Fig.~\ref{fig:spectra}.

Point B' is a point in the bulk region of the mSUGRA parameter space defined by the following
input parameters\footnote{Here we have used Isajet 7.69~\cite{Baer:2003mg} 
and DarkSUSY~\cite{Gondolo:2002tz} except for calculations where coannihilations are important, 
in which case we have used micrOMEGAS~\cite{Belanger:2001fz}.}: 
$m_0 = 57$ GeV, $m_{1/2} = 250$ GeV, $A_0 = 0$, $\tan \beta = 10$ and ${\rm sign}(\mu) = +1$.
As evident from Fig.~\ref{fig:spectra}, the particle spectrum at this point is quite light.  
The two lightest neuralinos, the lightest chargino and all of the sleptons have masses below $200$ GeV.  
All of the squarks are lighter than $600$ GeV.  The heaviest particle, the gluino, only weighs 
$611$ GeV. Therefore, one would expect colliders to have significant discovery and 
measurement capabilities.

Point LCC2 has been chosen in the focus point region~\cite{FocusPoint} of mSUGRA and has parameters
$m_0 = 3280$ GeV, $m_{1/2} = 300$ GeV, $A_0 = 0$, $\tan \beta = 10$ and ${\rm sign}(\mu) = +1$.
The masses of the squarks and sleptons are very heavy ($2-3$ TeV) while all charginos and neutralinos
are relatively light. At the LHC the dominant signal is expected to be due to gluino 
production~\cite{Azuelos:2002qw}. At the ILC500, all but the heaviest neutralino state 
can can be produced and the subsequent cascade decays to the LSP allow measurements of the SUSY
couplings and mass spectrum~\cite{Gray:2005ci}.

In Section~\ref{sec:Bp} we will use the expected precision of SUSY parameter measurements
for point B' at the LHC and ILC to derive the related uncertainty in $\Omega_\chi h^2$. 
The material in Section~\ref{sec:Bp} is based on Ref.~\cite{talk} (related studies have
later been performed in \cite{bulk}). In Section~\ref{sec:LCC2} we present the initial 
results of a similar, but more detailed analysis for the case of point LCC2.

\section{ANALYSIS FOR POINT B'}
\label{sec:Bp}

\begin{figure*}[t]
\centering
\includegraphics[width=55mm]{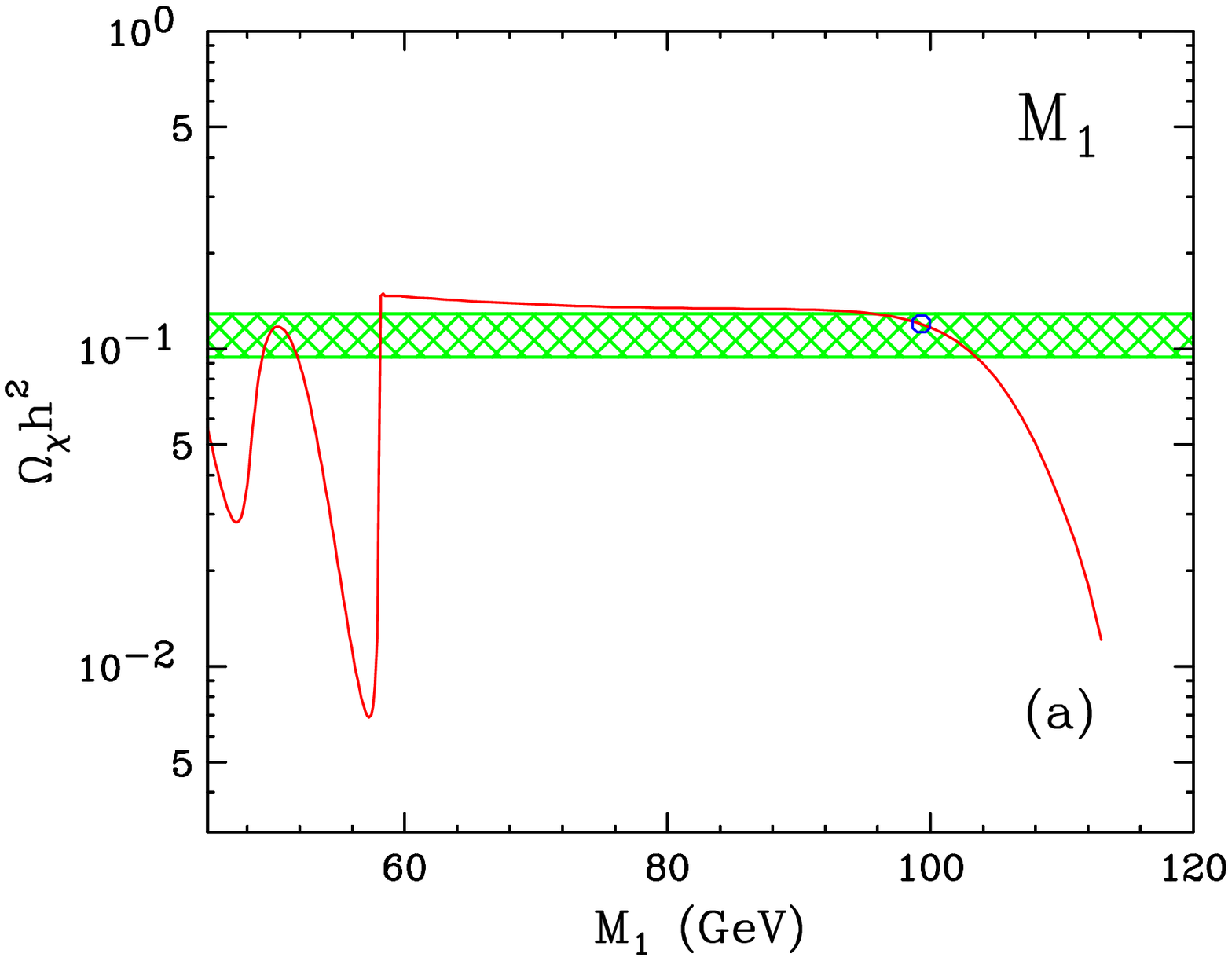}
\includegraphics[width=55mm]{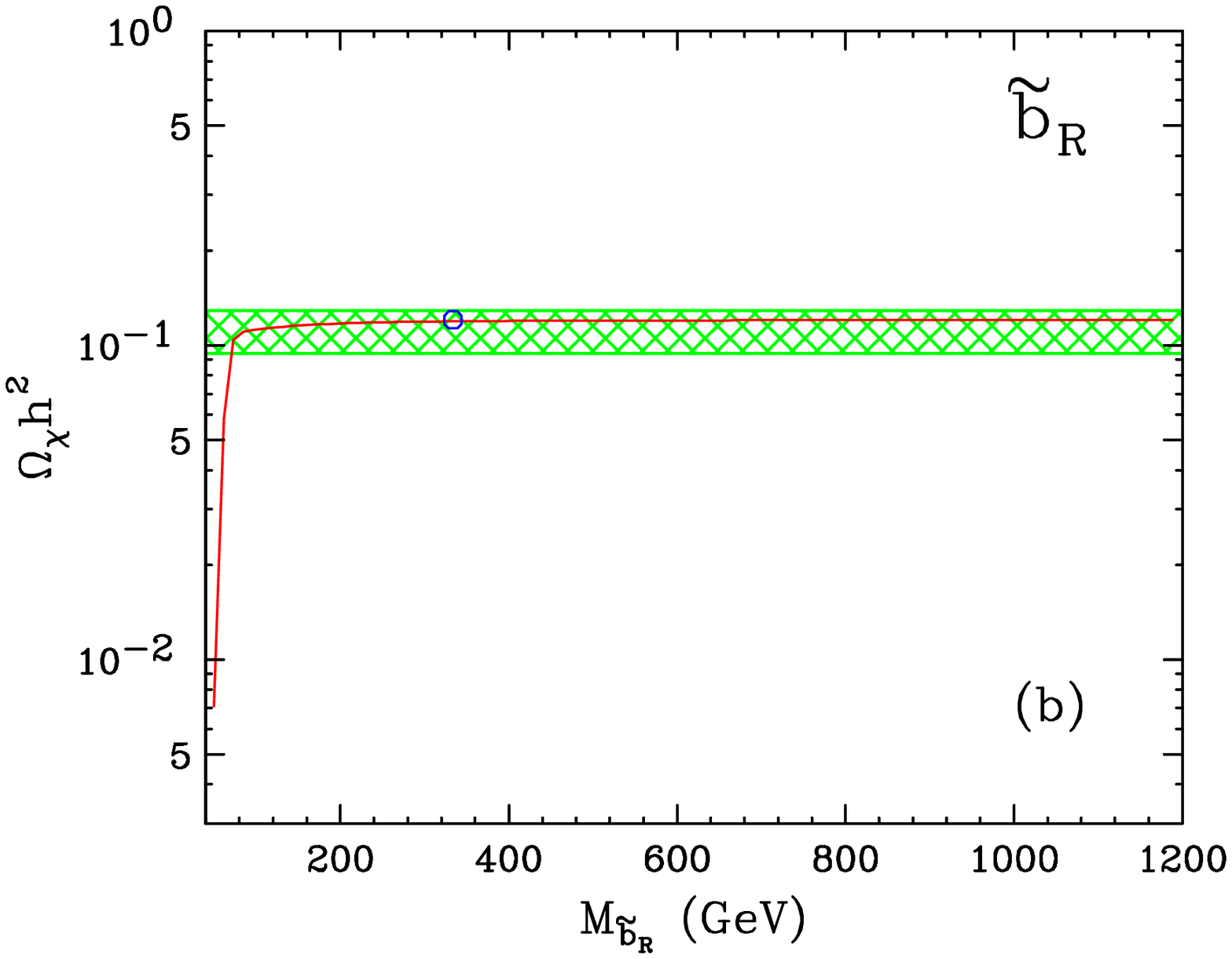}
\includegraphics[width=55mm]{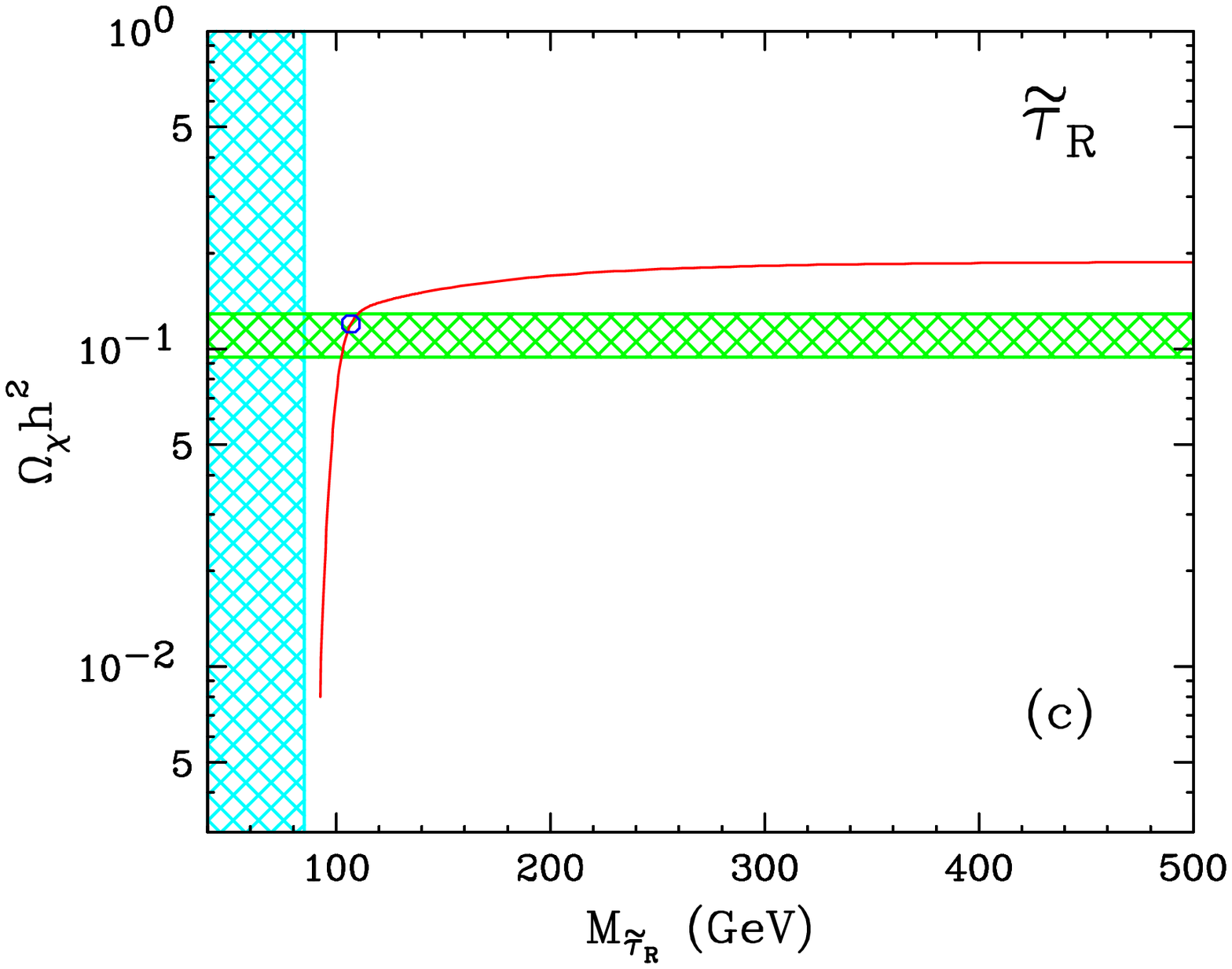}
\\
\centering
\includegraphics[width=55mm]{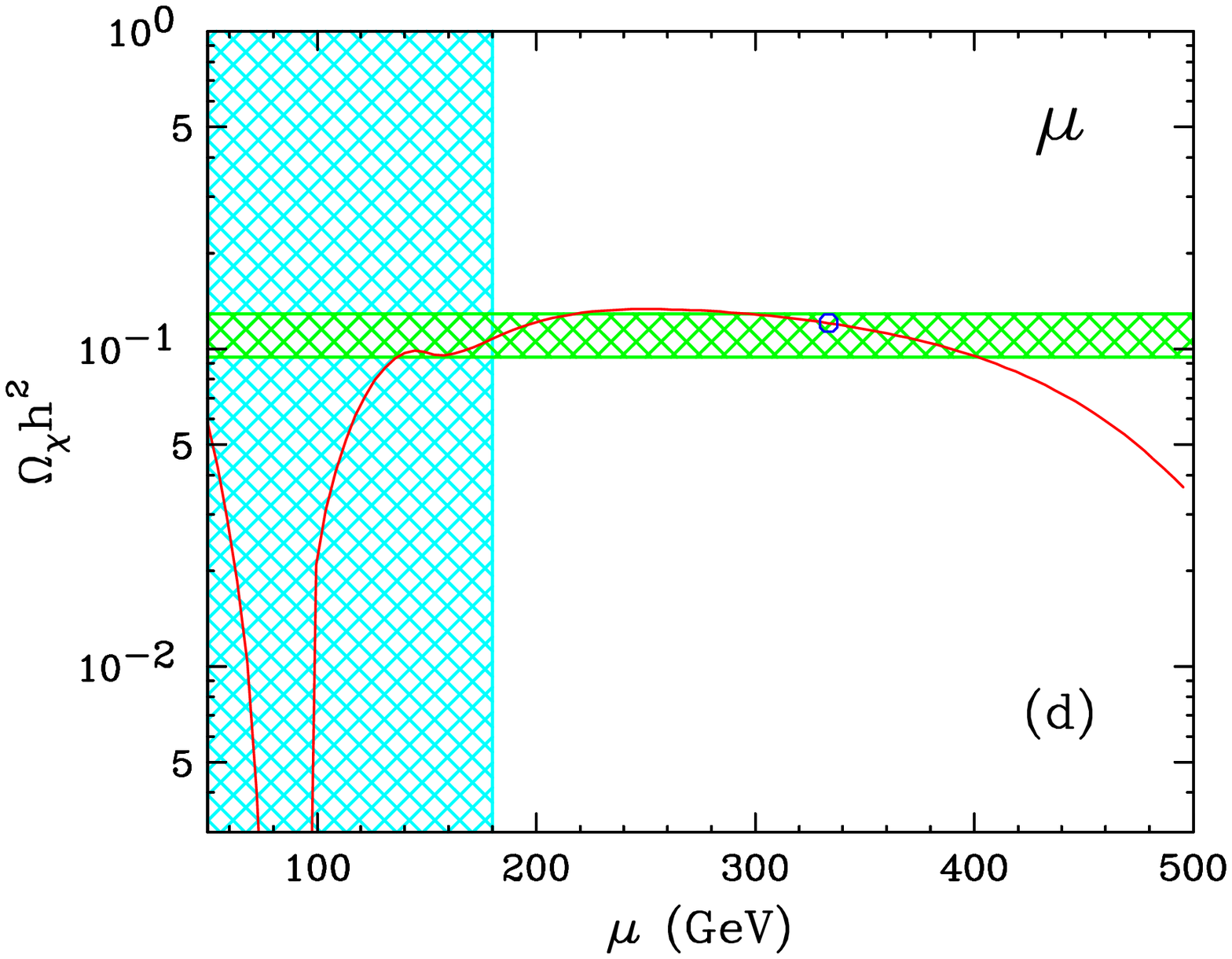}
\includegraphics[width=55mm]{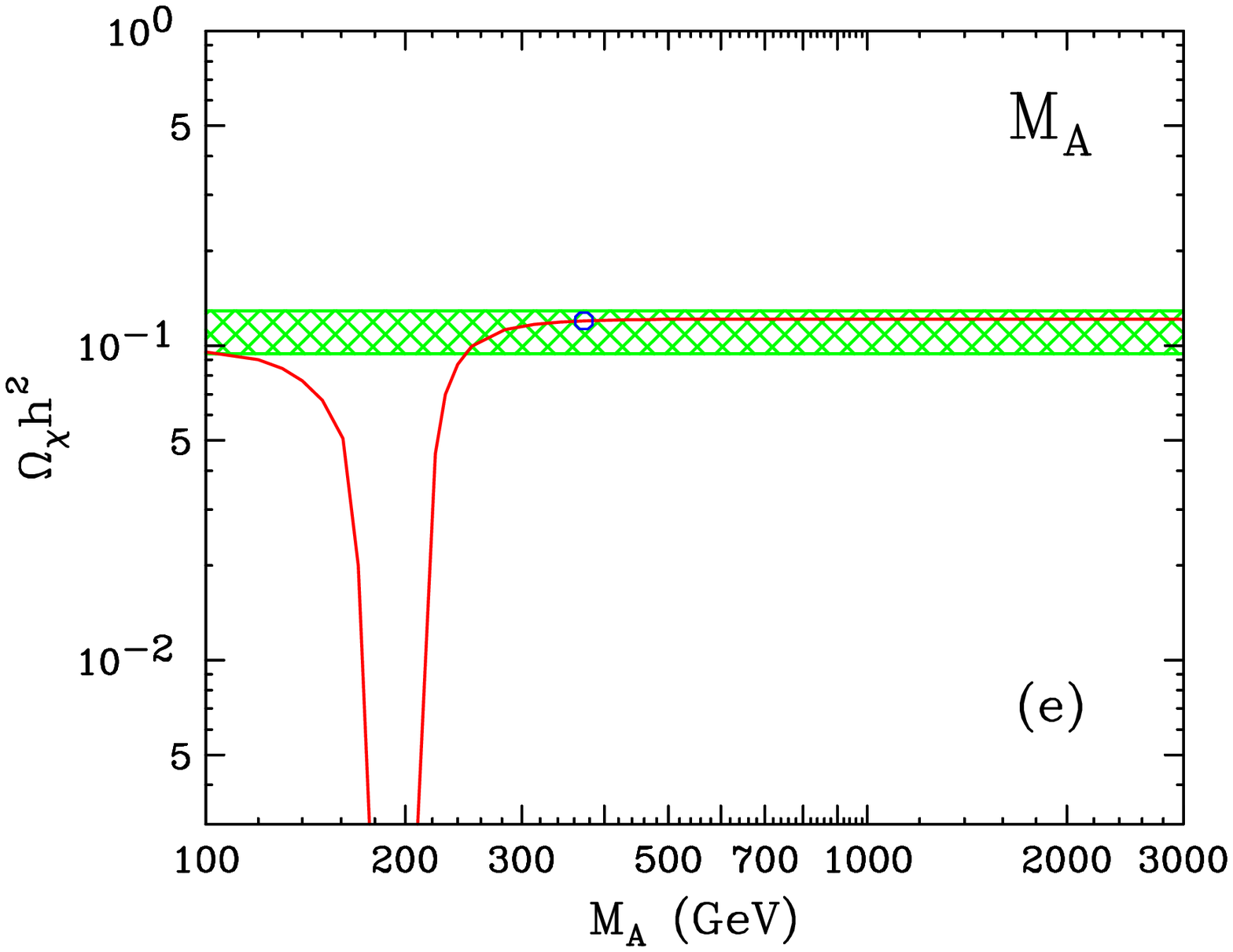}
\includegraphics[width=55mm]{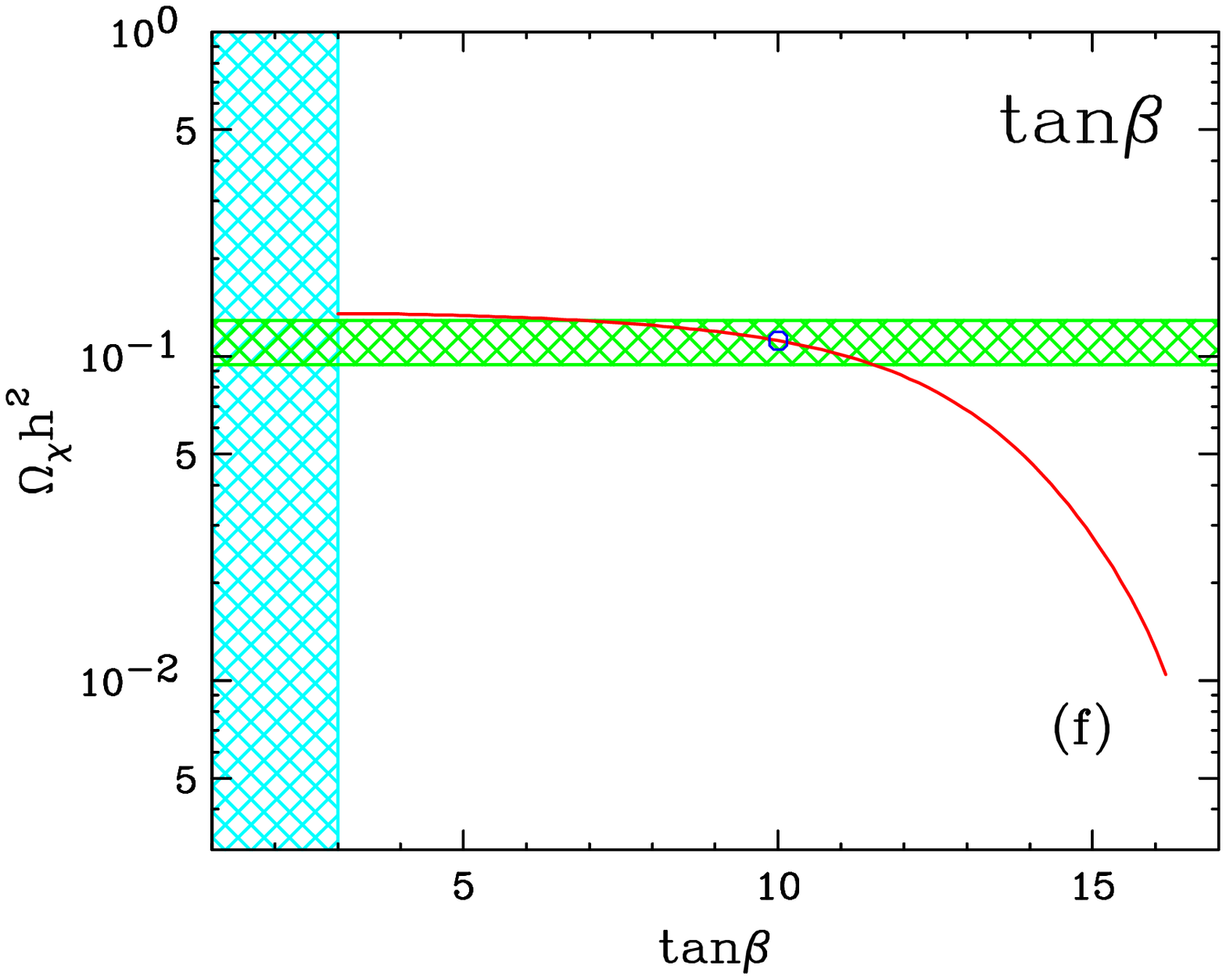}
\caption{Effect on relic density of varying relevant SUSY mass parameters for point B'.  
The horizontal (green-shaded) region denotes 
the $2\sigma$ WMAP limits on the dark matter relic density.  The red line shows 
the variation of the relic density as a function of the corresponding SUSY parameter. 
The vertical (blue-shaded) bands denote parameter regions currently ruled out by experiment.
The blue dot in each plot denotes the nominal value for the corresponding parameter at point B'.} 
\label{fig:Bpvary}
\end{figure*}

The analysis proceeds in two steps: first, we estimate the sensitivity of $\Omega_\chi h^2$
to the various SUSY parameters, and then we determine the precision with which they
can be measured at colliders. In Fig.~\ref{fig:Bpvary} we show the sensitivity of the
dark matter relic density to 6 relevant MSSM parameters. In each panel, the green region 
denotes the $2\sigma$ WMAP limits on $\Omega_\chi h^2$ and the red line shows the 
variation of the relic density as a function of the corresponding parameter.
The vertical (blue-shaded) bands denote parameter regions currently ruled out by experiment.
The blue dot in each panel denotes the nominal value for the corresponding parameter at point B'.

The importance of the different SUSY parameters for the determination of the neutralino 
relic density can be judged from the slope of the lines in Fig.~\ref{fig:Bpvary}. 
If $\Omega_\chi h^2$ is insensitive to a given parameter, the corresponding line 
will be flat: as we change the parameter, $\Omega_\chi h^2$ stays the same.
Conversely, if the relic density is particularly sensitiive to some SUSY parameter, the
slope of the corresponding variation curve will be very steep.
Fig.~\ref{fig:Bpvary} then reveals how different parameters affect $\Omega_\chi h^2$.
Interestingly enough, it also shows that even for one given parameter, there are regions 
where this parameter can be important (for example $M_{\tilde b_R}$ at low masses, where 
coannihilation processes with bottom squarks become important), as well as regions 
where this parameter has very little impact on the relic density (e.g.~$M_{\tilde b_R}$
at its nominal value). Therefore, in estimating the uncertainty on $\Omega_\chi h^2$,
collider data on otherwise ``irrelevant'' parameters can be very important.
In the absence of any information about the value of a given parameter, one should 
let it vary within the whole allowed range, which may encompass values of the
parameter for which it becomes relevant.

The behavior of the lines in Fig.~\ref{fig:Bpvary} can be understood as follows.
At point B' the lightest neutralino $\tilde\chi^0_1$ is mostly Bino, hence one 
would expect that its relic density will be sensitive to the Bino mass parameter $M_1$.
Indeed, this is confirmed by Fig.~\ref{fig:Bpvary}(a). For small $M_1$, we observe
enhanced sensitivity near the $Z$ and Higgs pole regions ($2M_{\tilde\chi}\sim M_Z$ 
and $2M_{\tilde\chi}\sim M_h$). For large values of $M_1$ we see a significant variation 
again, this time because the neutralino LSP becomes more and more degenerate with the 
sleptons, and its relic density is depleted due to coannihilation processes.
The analysis of Figs.~\ref{fig:Bpvary}(b) and \ref{fig:Bpvary}(c) is very similar:
the sfermions are irrelevant, if they are heavy, but may become very important if 
they are sufficiently light to induce coannihilations.
The result shown in Fig.~\ref{fig:Bpvary}(d) is somewhat complicated.
At low values of $\mu$ the LSP is pure higgsino, and its mass $M_\chi$ is determined 
by the higgsino mass parameter $\mu$. For $\mu$ between 90 and 100 GeV, we 
see the same $Z$ and Higgs pole regions which were evident in Fig.~\ref{fig:Bpvary}(a).
(The double dip structure is located in the range $10^{-4}<\Omega_\chi h^2 <10^{-3}$, 
which falls outside the plotted range). Notice, however, that $\mu<180$ GeV (the vertical 
(blue-shaded) band) implies a light higgsino-like chargino, which is ruled out by LEP. 
As $\mu$ gets larger, the LSP becomes Bino-like again, 
and its mass $M_\chi$ stops being dependent on $\mu$. This leads to a relatively
wide region of $\mu$ values around the nominal one, where $\mu$ is not very important.
However, at very large values of $\mu$ we see increased sensitivity again.
This is due to the effect of $\mu$ on stau mixing: as $\mu$ gets large, the 
off-diagonal components in the stau mass matrix increase as well, and push the smaller 
stau mass eigenvalue down, causing neutralino-stau coannihilations.
A similar effect is at play in Fig.~\ref{fig:Bpvary}(f), since the off-diagonal entries
in the stau mass matrix are proportional to $\tan\beta$ as well. 
Finally, Fig.~\ref{fig:Bpvary}(e) shows the sensitivity to the Higgs mass parameter
$M_A$, which controls the masses of the ``heavy''  Higgs bosons in the MSSM.
We see that apart from the Higgs pole around $M_A\sim 200$ GeV, where
$2 M_\chi\sim M_A$, the relic density is pretty much insensitive to $M_A$.

\begin{figure*}[t]
\centering
\includegraphics[width=80mm]{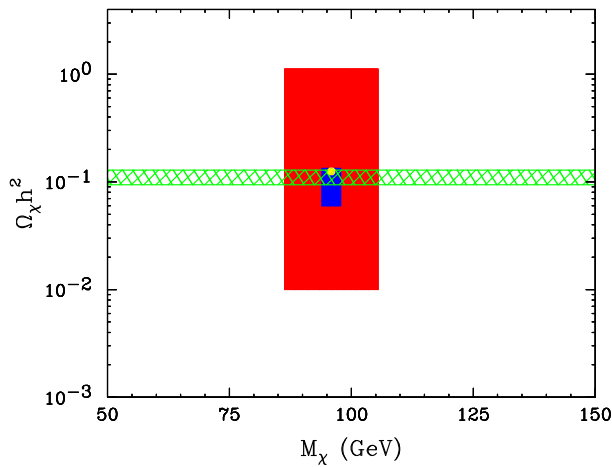}
\caption{Accuracy of WMAP (horizontal green shaded region), LHC (outer red rectangle) 
and ILC (inner blue rectangle) in determining $M_\chi$, the mass of the 
lightest neutralino, and its relic density $\Omega_\chi h^2$. 
The yellow dot denotes the actual values of $M_\chi$ and 
$\Omega_\chi h^2$ for point B'. } 
\label{fig:QUplot}
\end{figure*}

Having determined the correlations between the SUSY weak scale parameters
and the relic abundance of neutralinos, it is now straightforward to estimate
the uncertainty in $\Omega_\chi h^2$ after measurements at different colliders. 
The result is shown in Fig.~\ref{fig:QUplot}, where the outer red 
(inner blue) rectangle indicates the expected uncertainty at the LHC (ILC)
with respect to the mass $M_\chi$ and relic density $\Omega_\chi h^2$
of the lightest neutralino. The yellow dot denotes the actual 
values of $M_\chi$ and $\Omega_\chi h^2$ for point B' and the 
horizontal green shaded region is the current measurement (\ref{wmap}).

In arriving at this result, we made the following assumptions about the 
precision of the SUSY mass determinations at the LHC. We expect that the LHC will 
be able to detect gauginos in cascade decays of the left-handed squarks.
This may provide measurements of the $\tilde\chi^\pm_1$, $\tilde\chi^0_2$ and
$\tilde\chi^0_1$ masses at the level of 10\%. However, the remaining chargino 
and two neutralinos (i.e. the higgsinos) appear to be rather difficult to identify, 
which leads to a sizable uncertainty in the value of the $\mu$ parameter. 
Squark masses can be extracted by starting from events with gaugino decays 
and adding a jet to reconstruct the previous step up the decay chain.
The resulting precision should be no better than the precision on gaugino masses, 
but we have assumed 10\% again. The right-handed squarks are very challenging, 
as they lead to purely jetty signatures, and we assume we will have no first hand 
information on their spectrum. The sleptons present a challenge as well -- 
direct slepton production is plagued by large Standard Model backgrounds 
from $t\bar{t}$, $W^+W^-$ etc.~\cite{Andreev:2004qq} and we have assumed that 
sleptons cannot be directly observed. The right handed sleptons, however, are all
lighter than $\tilde\chi^0_2$, and may be produced in large quantities indirectly 
in gaugino cascade decays. Unfortunately, unless one is able to perform
a careful shape discrimination analysis~\cite{Birkedal:2005cm}, one
might easily confuse the sequential cascade $\tilde\chi^0_2\to\tilde\ell^\pm_R \ell^\mp
\to\tilde\chi^0_1 \ell^+\ell^-$ with that of heavy sleptons and direct
three-body decays $\tilde\chi^0_2\to\tilde\chi^0_1 \ell^+\ell^-$.
We have therefore conservatively assumed that no slepton information will be available.
Finally, in terms of Higgs bosons, we expect a detection only of the lightest 
(Standard Model-like) Higgs boson and the absence of a heavy Higgs boson 
signal will simply place the bound $M_A\geq200$ GeV. 

Our assumptions about the corresponding precision at ILC500 were the following.
Since superpartners need to be pair-produced, we take all sparticles lighter 
than 250 GeV to be observable, and their masses can be measured to within 2\%.
This includes the same chargino-neutralino states as in the case of LHC, plus 
all sleptons.

From Fig.~\ref{fig:QUplot} we see that with the assumptions above, the LHC (scheduled to 
turn on in 2007) is not competitive with the current state of the art 
determination of the relic density from cosmology. Nevertheless, it will
bound the relic density from above and below, and may provide the first hint
on whether the dark matter candidate being discovered at colliders is 
indeed the dark matter of cosmology. The ILC will fare much better, and will 
achieve a precision rivalling that of the cosmological determinations.

\section{ANALYSIS FOR POINT LCC2}
\label{sec:LCC2}

\begin{figure*}[t!]
\centering
\includegraphics[width=60mm]{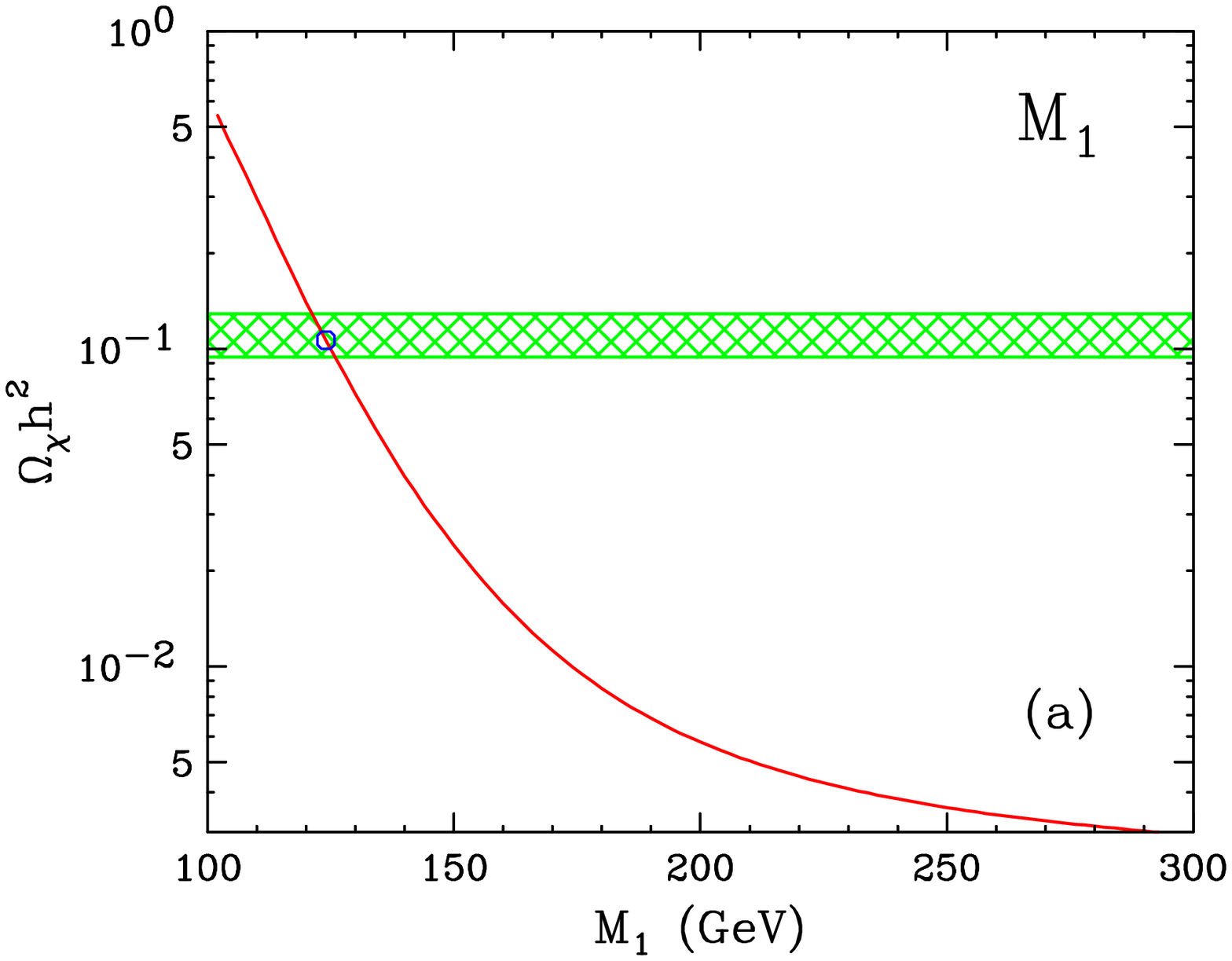}
\includegraphics[width=60mm]{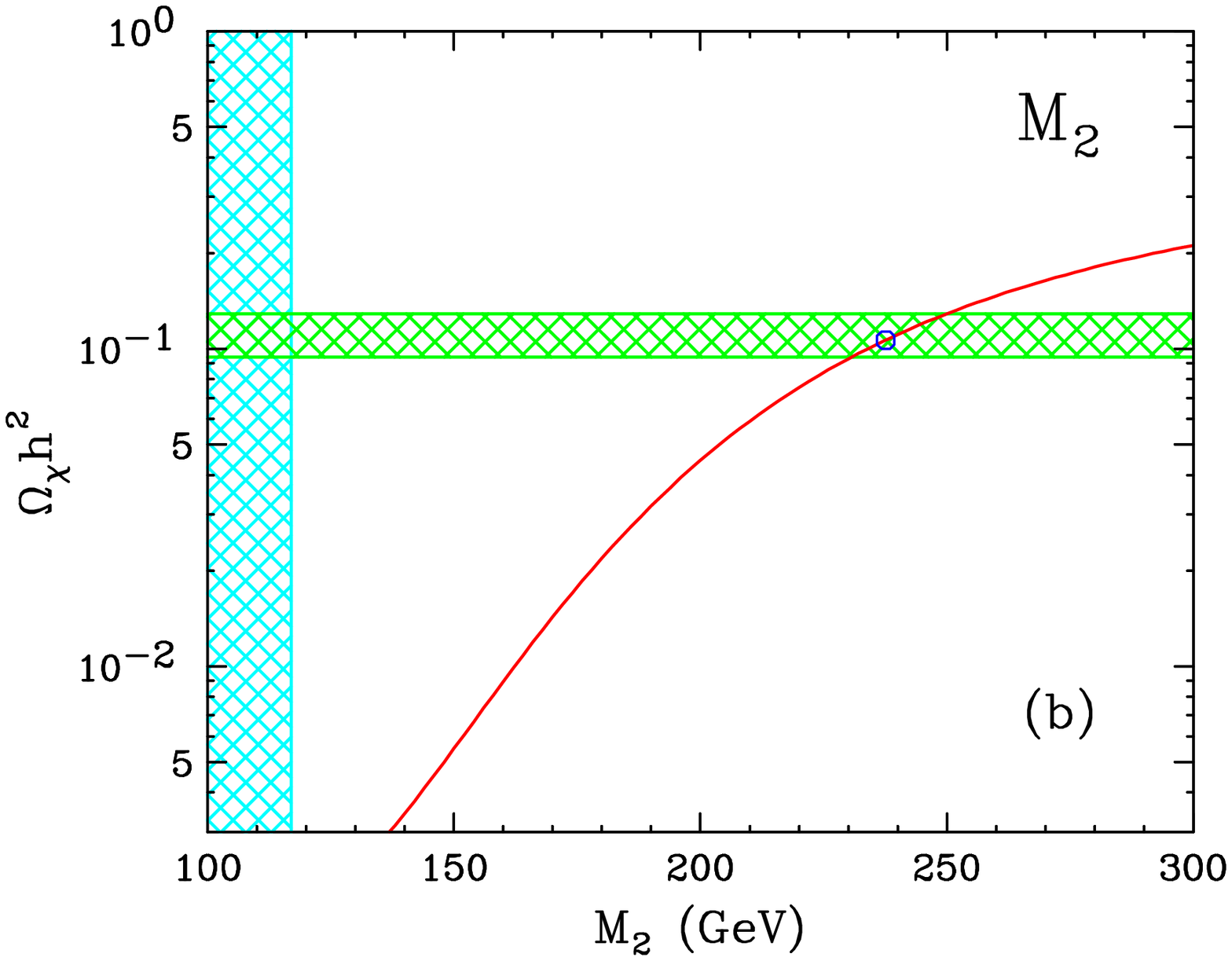}
\\
\includegraphics[width=60mm]{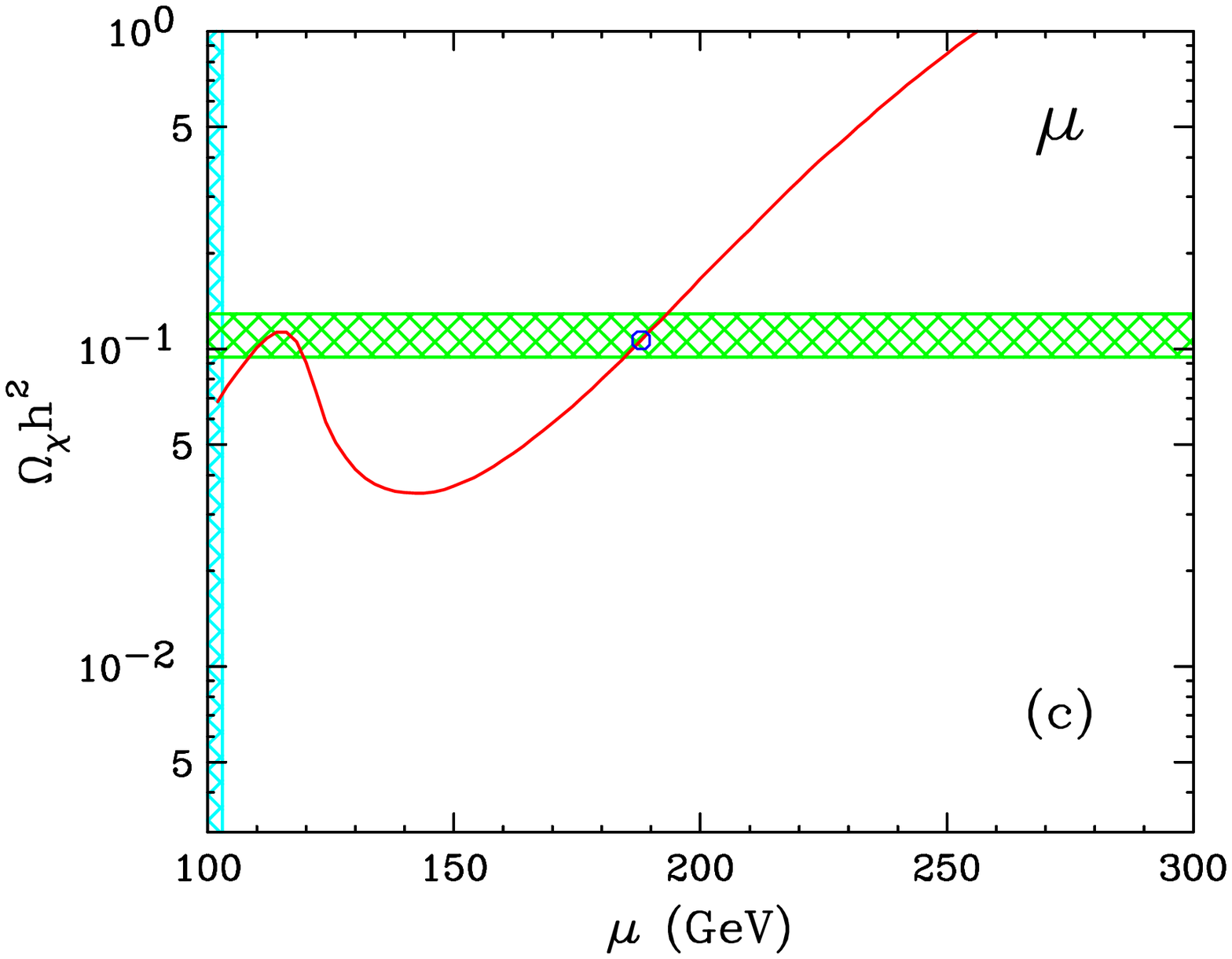}
\includegraphics[width=60mm]{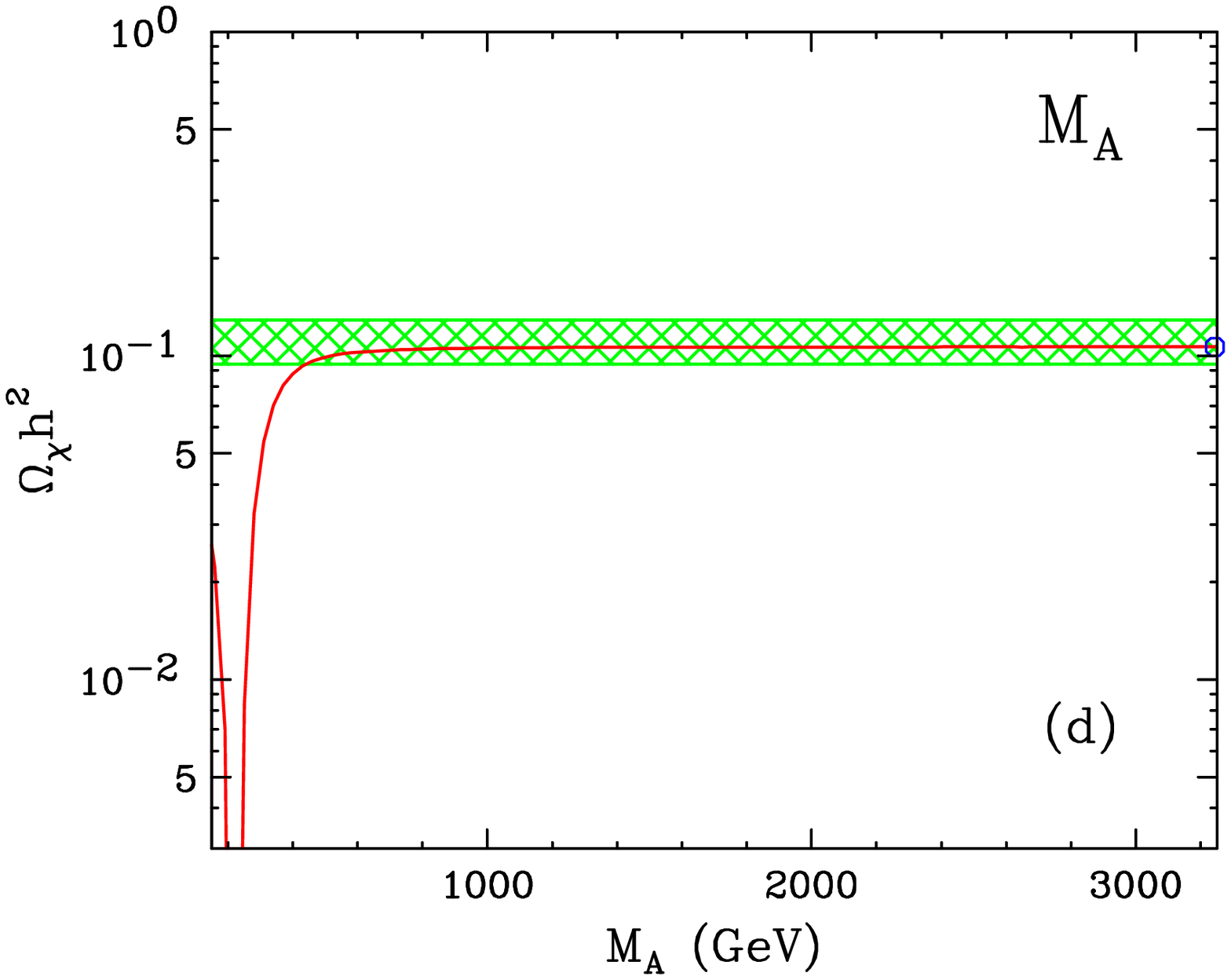}

\caption{The same as Fig.~\ref{fig:Bpvary}, but for the point LCC2.} 
\label{LCC2vary}
\end{figure*}
Turning to point LCC2, we show in Fig.~\ref{LCC2vary} 
the analogous variation of the relic density as a function of 4 relevant parameters.
Here the squark and slepton masses are heavy and have little impact on 
the actual $\Omega_\chi h^2$. For this point, the lightest neutralino is mostly 
Bino, but with a non-negligible higgsino component. 

Fig.~\ref{LCC2vary}(a) shows the dependence of $\Omega_\chi h^2$
on the Bino mass parameter $M_1$. As expected, lowering $M_1$ 
increases the Bino component of the LSP, thus suppressing 
$\sigma_{\rm an}$ and increasing $\Omega_\chi h^2$. 
Fig.~\ref{LCC2vary}(c) exhibits complementary behavior: 
lowering $\mu$ increases the higgsino component, 
enhancing $\sigma_{\rm an}$ and lowering  $\Omega_\chi h^2$.
Fig.~\ref{LCC2vary}(b) is in a sense similar to
Fig.~\ref{LCC2vary}(c): the $M_2$  parameter controls the wino fraction 
of the LSP, and small values of $M_2$ lead to wino-like dark matter, which 
has a large annihilation rate and therefore smaller relic abundance\footnote{Neutralino dark matter with significant wino content has been studied in the contexts of scenarios with non-universal gaugino masses~\cite{NonUniversal}, string-derived supergravity~\cite{StringDerived}, and anomaly mediated supersymmetry breaking~\cite{AnomalyMediation}.}. 
Finally, Fig.~\ref{LCC2vary}(d) is the analog of
Fig.~\ref{fig:Bpvary}(e) for the case of point LCC2.

These results will be combined with the outcome of a comprehensive simulation study,
including detailed detector simulation, on the expected experimental precision 
at the ILC500 for point LCC2. The final product will be the analog of 
Fig.~\ref{fig:QUplot}. For further details on the current status of the analysis 
for point LCC2, see~\cite{Gray:2005ci}.

\begin{acknowledgments}
The work of AB and KM is supported by a US Department of Energy
Outstanding Junior Investigator award under grant DE-FG02-97ER41029.
\end{acknowledgments}


\end{document}